\documentclass[11pt]{article}
\usepackage{array}
\usepackage{multirow}

\usepackage{graphics}
\usepackage[dvips]{graphicx}
\usepackage[a4paper,left=1.4cm,right=1.5cm,top=1.5cm]{geometry}
\usepackage{subcaption} 

\begin{document}

\pagestyle{myheadings}
\markright{\footnotesize {doi:10.6062/jcis.2016.07.03.0114 \hspace{3,2cm} Discussion Paper \hspace{4,3cm} Barchi et al.}}  

\vspace{1.2cm}

\begin{center}

{\bf  {\Large Improving galaxy morphology with machine learning}}
\bigskip


{\small P. H. Barchi$^a$\footnote{E-mail Corresponding Author: paulobarchi@gmail.com}, R. Sautter$^a$, 
F. G. da Costa$^b$, T. C. Moura$^a$, D. H. Stalder$^a$, R.R. Rosa$^a$ and R.R. de Carvalho$^a$ 
}
\smallskip

{\small
$^a$National Institute for Space Research (INPE), S\~ao Jos\'e dos Campos, SP, Brazil\\

$^b$University of S\~ao Paulo (USP), S\~ao Carlos, SP, Brazil\\
}

{\footnotesize Received on November 15, 2016 / accepted on December 30, 2016}

\end{center}

\quad


\begin{abstract}
This paper presents machine learning experiments performed over results of galaxy 
classification into elliptical (E) and spiral (S) with morphological parameters: concentration 
(CN), assimetry metrics (A3), smoothness metrics (S3), entropy (H) and gradient pattern 
analysis parameter (GA). Except concentration, all parameters performed a image segmentation 
pre-processing. For supervision and to compute confusion matrices, we used as true label the 
galaxy classification from GalaxyZoo. With a 48145 objects dataset after preprocessing 
(44760 galaxies labeled as S and 3385 as E), we performed experiments with Support Vector 
Machine (SVM) and Decision Tree (DT). Whit a 1962 objects balanced dataset, we applied K-means 
and Agglomerative Hierarchical Clustering. All experiments with supervision reached an 
Overall Accuracy $OA \geq 97\%$. 

\quad

{\footnotesize
{\bf Keywords}: Machine Learning, Computational Astrophysics.}
\end{abstract}


\quad

\textbf{1. Introduction}
\smallskip

The volume of digital data of stars, galaxies, and the universe has multiplied in 
recent decades due to the rapid development of new technologies as new satellites, 
telescopes, and other observatory instruments. The process of scientific discovery 
is increasingly dependent on the ability to analyse massive amounts of complex data 
from scientific instruments and simulations. Such analysis has become the 
bottleneck of the scientific process 
\cite{efficientMLreview,advancesML4Astro,statisticsAstroML,statChallengesInAstro,dataMining}.

By studying global properties of early-type galaxies (ETGs), 
researchers from the thematic project \cite{dedrives} have been able to 
constrain models of galaxy formation and evolution. These thematic project in 
progress \cite{dedrives,special} extends galaxy evolution studies to consistently 
investigate galaxies and their environments over a significant time baseline. 
To this end, Ferrari et al. \cite{morfometryka} presents an extended morphometric system to 
automatically classify galaxies from astronomical images and Andrade et al. \cite{gpa} 
introduces the preliminary results of the characterization of pattern 
evolution in the process of cosmic structure formation.

In the context of these projects, this paper presents the first steps towards 
improving galaxy morphology with Machine Learning (ML). The dataset used for
supervised learning experiments consists of 48145 objects after preprocessing, 
with 44760 galaxies labeled as S and 3385 as E. The preprocessing removed 3611 
objects with missing data for one of the features: CN. We used as features of 
the dataset the best morphological parameters from each type to classify 
galaxies: concentration (CN), assimetry metrics (A3), smoothness metrics (S3), 
entropy (H) and gradient pattern analysis parameter (sGA). These are preliminary 
results from an ongoing research about morphological parameters to classify 
galaxies into spiral (S) and elliptical (E) -- a full publication about it will 
be released soon \cite{Sautter}. The target of our dataset (considered as true 
label) is the classification from Galaxy Zoo project \cite{galaxyZoo}. 
The experiments were conducted to explore different method parametrization, 
if it is applicable. For the unsupervised learning experiments, we used
a balanced dataset with 1962 objects.

As related work of ML in this astrophysical context, 
Ball and Brunner \cite{ball2010} surveys a long list of data mining
and ML projects for analyzing astronomical data. Ivezić
et al. \cite{statisticsAstroML} provides modern statistical methods for 
analyzing astronomical data. Vasconcellos et al. \cite{vasconcellos} employ 
decision tree classifiers for star/galaxy separation. And more recently, 
Schawinski et al. \cite{GAN-galaxies} used Generative Adversarial Networks 
(GAN) to recover features in astrophysical images of galaxies.

In Sections 2 and 3 we present the experiments performed with supervised 
and unsupervised ML methods, respectivelly, using scikit-learn 
python library \cite{scikit}. In all experiments we explored the 5 parameters 
with best confusion matrices obtained by Sautter \cite{Sautter} for galaxy 
classification: CN, A3, S3, H, and Ga. In this work, the confusion matrices 
for each experiment present the necessary values to calculate the metrics 
presented in Tables \ref{tab:comp1} and \ref{tab:comp2} -- for each class 
(S and E): true positives (TP - correctly classifyed objects), false 
positives (FP - error, galaxies which are not from this class and classifyed 
as such), true negatives (TN - objects correctly rejected in classification 
for this class), and false negatives (FN - error, galaxies mistankenly 
rejected to be classified for such class). We conclude the paper presenting 
the results and final considerations in Section 4.

\quad

\bigskip

\textbf{2. Supervised Learning Methods}
\smallskip

Supervised Learning (SL) is a learning process guided by some form of supervision 
to build a model to perform the approached task. This supervision may be associated, 
for example, with a previously labeled sample; from these, patterns can be identified 
to sort or group new, unlabelled examples. For this, the dataset must be split
into different sets to train, validate and test the model. 

There are various approaches conserning the split of the whole dataset into trainning,
validating and test sets for supervised methods. By partitioning the available data into 
three sets, we drastically reduce the number of samples which can be used for learning the 
model, and the results can depend on a particular random choice for the pair of (train, 
validation) sets. A solution to this problem is a procedure called cross-validation (CV):
a way to address the tradeoff between bias and variance. In the basic CV approach used in 
this work, denominated k-fold CV, a model is trainned using k-1 of the folds as trainning 
data \cite{scikit}. 

So, for these experiments, first we split the dataset, for instance, in a 80-20 proportion 
for tranning and test sets, respectively. From the 80\% of the first split, we apply k-fold 
CV for the trainning phase, with k = 20 folds and k = 5 folds. The resulting model is 
validated on the remaining part of the data. With this procedure, the model is ready to the 
test phase. Then, we test with the 20\% remaining data from the first split. Analogously, 
we also made experiments with a 50-50 proportion for the first split. As mentioned in
Section 1, the dataset used for these experiments consists of 48145 objects after 
preprocessing, with 44760 galaxies labeled as S and 3385 as E.

We also used a Grid Search (GS) to exhaustively generate candidates from a grid of parameter 
values. For the case of the Decision Tree (DT) classifiers, the parameter values are relative 
to the depth of the DT. When fitting the model to the dataset, all possible combinations of 
parameter values are evaluated and the best combination is retained. This is done using the CV 
score which is basically a convenience wrapper for the sklearn cross-validation iterators. 
Given a classifier (such as Support Vector Machine) and the dataset for trainning phase (in this 
case, 80\% of the whole dataset for trainning and validating), it automatically performs 
rounds of CV by splitting trainning/validation sets, fitting the trainning and computing 
the score on the validation set. GS and CV score used here are provided by 
\textit{GridSearchCV} and \textit{cross\_val\_score} from scikit-learn python library 
\cite{scikit}.

\quad

\textit{\textbf{2.1. Support Vector Machines (SVM)}}
\smallskip

On the problem of binary classification, it is possible to draw infinite different
hyperplanes for separating both classes so that the error rate reaches a minimum.
Support Vector Machines (SVM) constructs the optimal hyperplane that will divide 
the target classes. An optimal hyperplane is the one that maximizes the separation
margins between the classes, providing a unique solution for the problem \cite{svm}.

When the input data is not linearly separable, i.e., the input space can not be
separated by a line, the Support Vector Machines implement the 'kernel-trick',
in which the input space is mapped into some high dimensional feature space
through some non-linear mapping chosen a priori. This mapping is done by a 
dot-product in the feature space, by an N-dimensional vector function $\phi(\cdot)$,
which can be a polynomial function, a radial basis function or other \cite{svm2}.

We conducted 4 experiments with SVM described below. The Table \ref{tb:svm} 
presents their confusion matrices.
\begin{itemize}
 \item \#SVM1: K-fold CV with k = 5 and data split in 50-50 proportion for trainning and test;
 \item \#SVM2: K-fold CV with k = 5 and data split in 80-20 proportion for trainning and test;
 \item \#SVM3: K-fold CV with k = 20 and data split in 50-50 proportion for trainning and test;
 \item \#SVM4: K-fold CV with k = 20 and data split in 80-20 proportion for trainning and test;
\end{itemize}

\begin{table}[!ht] 
  \centering
  \setlength{\extrarowheight}{2pt}
  \begin{tabular}{cc|c|c|}
    & \multicolumn{1}{c}{} & \multicolumn{2}{c}{Pred. label}\\
    & \multicolumn{1}{c}{} & \multicolumn{1}{c}{S}  & \multicolumn{1}{c}{E} \\\cline{3-4}
    True  & S & 22044 & 323 \\\cline{3-4}
    label & E & 294 & 1412 \\\cline{3-4}
  \end{tabular}
  \quad
  \setlength{\extrarowheight}{2pt}
  \begin{tabular}{cc|c|c|}
    & \multicolumn{1}{c}{} & \multicolumn{2}{c}{Pred. label}\\
    & \multicolumn{1}{c}{} & \multicolumn{1}{c}{S}  & \multicolumn{1}{c}{E} \\\cline{3-4}
    True  & S & 8807 & 125 \\\cline{3-4}
    label & E & 135 & 562 \\\cline{3-4}
  \end{tabular}
  \\
  \setlength{\extrarowheight}{2pt}
  \begin{tabular}{cc|c|c|}
    & \multicolumn{1}{c}{} & \multicolumn{2}{c}{Pred. label}\\
    & \multicolumn{1}{c}{} & \multicolumn{1}{c}{S}  & \multicolumn{1}{c}{E} \\\cline{3-4}
    True  & S & 22092 & 280 \\\cline{3-4}
    label & E & 347 & 1354 \\\cline{3-4}
  \end{tabular}
  \quad
  \setlength{\extrarowheight}{2pt}
  \begin{tabular}{cc|c|c|}
    & \multicolumn{1}{c}{} & \multicolumn{2}{c}{Pred. label}\\
    & \multicolumn{1}{c}{} & \multicolumn{1}{c}{S}  & \multicolumn{1}{c}{E} \\\cline{3-4}
    True  & S & 8861 & 106 \\\cline{3-4}
    label & E & 132 & 530 \\\cline{3-4}
  \end{tabular}
  \caption{Confusion Matrices for the experiments with Support Vector Machine (SVM1, SVM2, 
  SVM3 and SVM4, from left to right, respectively).}
  \label{tb:svm}
\end{table}

\quad

\textit{\textbf{2.2. Decision Tree (DT)}}
\smallskip

Decision Tree (DT) is a supervised machine learning method to classification and
regression. The goal here is to create a model which predicts the classification by
learning simple decision rules inferred from the dataset \cite{tree}. 
Classification and Regression Tree (CART) is very similar to the C4.5 Decision
Tree algorithm, but it supports numerical target values and
does not compute rule sets. CART builds binary trees using feature and threshold
that yields the largest information gain at each node. We used the optimized 
version of CART algorithm provided by scikit-learn python library \cite{scikit}.

The experiments with DT followed the same procedure performed with SVM, and their 
confusion matrices are shown in Table \ref{tb:dt}.
\begin{itemize}
 \item \#DT1: K-fold CV with k = 5 and data split in 50-50 proportion for trainning and test;
 \item \#DT2: K-fold CV with k = 5 and data split in 80-20 proportion for trainning and test;
 \item \#DT3: K-fold CV with k = 20 and data split in 50-50 proportion for trainning and test;
 \item \#DT4: K-fold CV with k = 20 and data split in 80-20 proportion for trainning and test;
\end{itemize}


\begin{table}[!ht]
  \centering
  \setlength{\extrarowheight}{2pt}
  \begin{tabular}{cc|c|c|}
    & \multicolumn{1}{c}{} & \multicolumn{2}{c}{Pred. label}\\
    & \multicolumn{1}{c}{} & \multicolumn{1}{c}{S}  & \multicolumn{1}{c}{E} \\\cline{3-4}
    True  & S & 22167 & 255 \\\cline{3-4}
    label & E & 271 & 1380 \\\cline{3-4}
  \end{tabular}
  \quad
  \setlength{\extrarowheight}{2pt}
  \begin{tabular}{cc|c|c|}
    & \multicolumn{1}{c}{} & \multicolumn{2}{c}{Pred. label}\\
    & \multicolumn{1}{c}{} & \multicolumn{1}{c}{S}  & \multicolumn{1}{c}{E} \\\cline{3-4}
    True  & S & 8841 & 81 \\\cline{3-4}
    label & E & 160 & 547 \\\cline{3-4}
  \end{tabular}
  \\
  \setlength{\extrarowheight}{2pt}
  \begin{tabular}{cc|c|c|}
    & \multicolumn{1}{c}{} & \multicolumn{2}{c}{Pred. label}\\
    & \multicolumn{1}{c}{} & \multicolumn{1}{c}{S}  & \multicolumn{1}{c}{E} \\\cline{3-4}
    True  & S & 22089 & 295 \\\cline{3-4}
    label & E & 297 & 1392 \\\cline{3-4}
  \end{tabular}
  \quad
  \setlength{\extrarowheight}{2pt}
  \begin{tabular}{cc|c|c|}
    & \multicolumn{1}{c}{} & \multicolumn{2}{c}{Pred. label}\\
    & \multicolumn{1}{c}{} & \multicolumn{1}{c}{S}  & \multicolumn{1}{c}{E} \\\cline{3-4}
    True  & S & 8827 & 111 \\\cline{3-4}
    label & E & 116 & 575 \\\cline{3-4}
  \end{tabular}
  \caption{Confusion Matrices for the experiments with Decision Tree (DT1, DT2, DT3 and DT4, 
  from left to right, respectively).}
  \label{tb:dt}
\end{table}
\quad

%
\quad

\textbf{3. Unsupervised Learning Methods}
\smallskip

Unsupervised Learning (UL), as can be inferred from the name, differs from SL because 
has no supervision, i.e., there is no model to guide the learning process.
One of the most common tasks in UL is to form groups of non-labeled examples
according to their similarities, process also denominated as clustering. 

In this unsupervised context, we used no supervision at all for clustering methods 
to obtain the perfomance of each method with a balanced dataset with 1962 objects,
without considering the GalaxyZoo labels, aiming to prepare for future unlabeled datasets.

\quad

\textit{\textbf{3.1. K-means Clustering}}
\smallskip

One of the more general-purpose clustering methods (non-supervised machine learning), 
K-means finds clusters of similar sizes, flat geometry, not many clusters, and accepts 
specification of clusters \cite{kmeans}. Among the possible variations of this clustering 
algorithm in scikit-learng library, it is possible to vary the number of times that the 
algorithm will execute with different seeds as centroids; However, with tests varying this 
value between 10, 100, 1000, there was no relevant variation in the resulting cluster. 
Another possible variation is related to the method to start the selection of the centers 
of the algorithms: `k-means ++' accelerates the convergence and `random' selects randomly. 
With this clustering method, we conducted 2 experiments, one using 'k-means++' 
(\emph{K-means1}) and the other 'random' (\emph{K-means2}), which both obtained the same result.

\begin{table}[!ht]
  \centering
  \setlength{\extrarowheight}{2pt}
  \begin{tabular}{cc|c|c|}
    & \multicolumn{1}{c}{} & \multicolumn{2}{c}{Pred. label}\\
    & \multicolumn{1}{c}{} & \multicolumn{1}{c}{S}  & \multicolumn{1}{c}{E} \\\cline{3-4}
    True  & S & 831 & 137 \\\cline{3-4}
    label & E & 63 & 931 \\\cline{3-4}
  \end{tabular}
  \quad
  \setlength{\extrarowheight}{2pt}
  \begin{tabular}{cc|c|c|}
    & \multicolumn{1}{c}{} & \multicolumn{2}{c}{Pred. label}\\
    & \multicolumn{1}{c}{} & \multicolumn{1}{c}{S}  & \multicolumn{1}{c}{E} \\\cline{3-4}
    True  & S & 831 & 137 \\\cline{3-4}
    label & E & 63 & 931 \\\cline{3-4}
  \end{tabular}
  \caption{Confusion Matrices for the experiments with K-means Clustering.}
\end{table}

\quad

\textit{\textbf{3.2. Agglomerative Hierarchical Clustering (AHC)}}
\smallskip

Starting with all $n$ objects to be clustered, AHC groups these objects into successively 
fewer than $n$ sets. It is a hierarchical nonoverlapping method that specify a sequence 
$P_0, ..., P_w$ of partitions of the objects in which $P_0$ is the disjoint partition, 
$P_w$ is the conjoint partition, and $P_i$ is a refinement (in the usual sense) of 
$P_j$ for all $0 \leq i < j \leq w$. It is a sequential method since the same algorithm
is used iteratively to generate $P_{i+1}$ from $P_i$ for all $0 \leq i < w$. Is is a 
pair-group method: at each iteration exactly two clusters are agglomerated into a single 
cluster \cite{ahc}. Although it is more suitable to find many clusters through this
unsupervised approach, in this experiment we used Agglomerative Hierarchical Clustering 
(AHC) to find two clusters (\emph{AHC} experiment), i.e., our result is the two main
subgroups obtained from the whole datasets (the smaller subgroups are irrelevant for 
this experiment).

\begin{figure}[!ht]
  \centering
  \begin{subfigure}[t]{0.5\textwidth}
    \includegraphics[width=\textwidth]{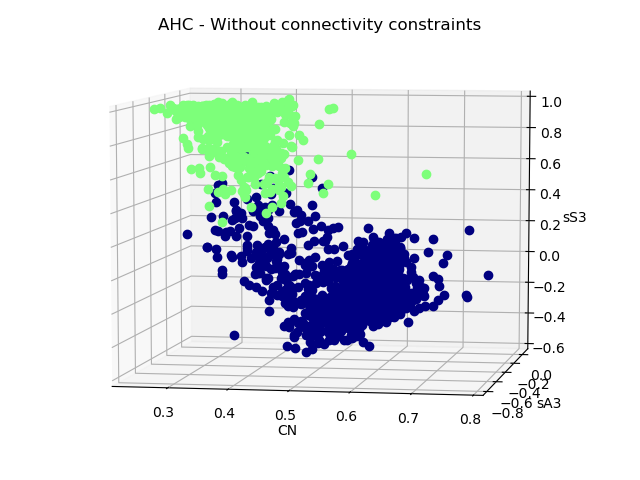}
  \end{subfigure}%
  ~ 
  \begin{subfigure}[t]{0.5\textwidth}
    \includegraphics[width=\textwidth]{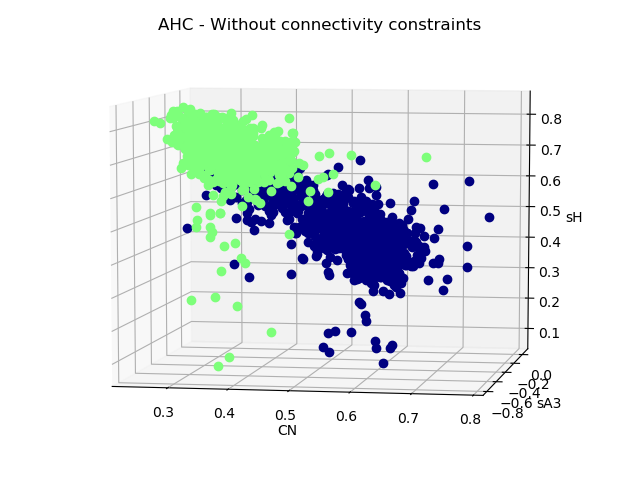}
  \end{subfigure}
  \\
  \begin{subfigure}[t]{0.5\textwidth}
    \includegraphics[width=\textwidth]{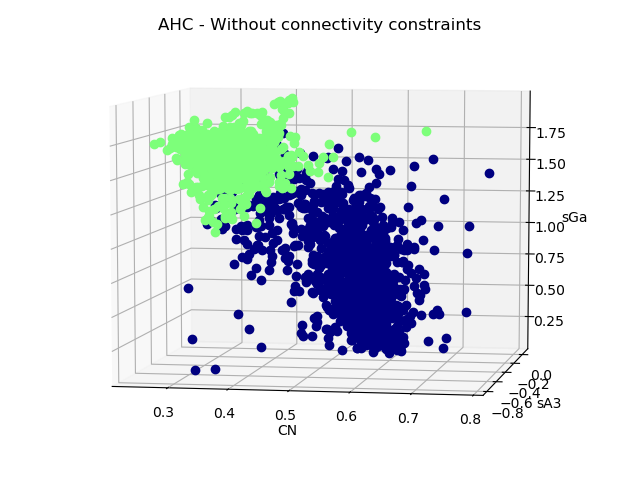}
  \end{subfigure}%
  ~
  \begin{subfigure}[t]{0.5\textwidth}
    \includegraphics[width=\textwidth]{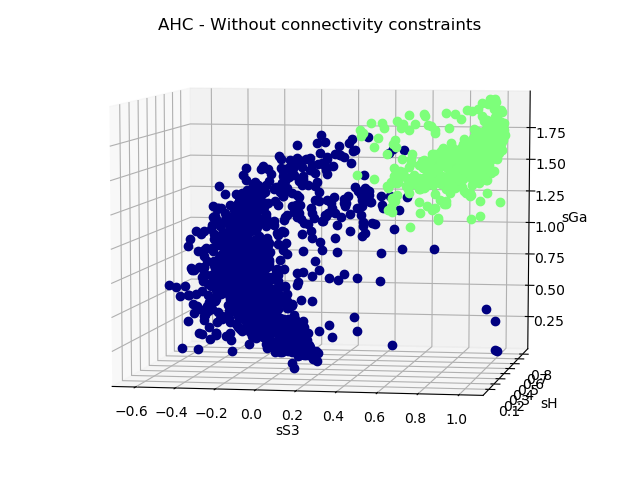}
  \end{subfigure}
  \caption{Reprentations in three dimensions of the result of Agglomerative Hierarchical Clustering.}
  \label{fig:AHC}
\end{figure}

The resulting Agglomerative Hierarchical Clustering with the two clusters found is 
represented in Figure \ref{fig:AHC} with all possible three dimensional representations, 
i.e., three morphological parameters combined at time to build the 3D data space.

\begin{table}[ht]
  \centering
  \setlength{\extrarowheight}{2pt}
  \begin{tabular}{cc|c|c|}
    & \multicolumn{1}{c}{} & \multicolumn{2}{c}{Pred. label}\\
    & \multicolumn{1}{c}{} & \multicolumn{1}{c}{S}  & \multicolumn{1}{c}{E} \\\cline{3-4}
    True  & S & 946 & 48 \\\cline{3-4}
    label & E & 175 & 793 \\\cline{3-4}
  \end{tabular}
  \caption{Confusion Matrix for the experiment with Agglomerative Hierarchical Clustering.}
\end{table}

\clearpage

\textbf{4. Concluding Remarks}
\smallskip

The Tables \ref{tab:comp1} and \ref{tab:comp2} present a comparative summary 
of the supervised and unsupervised experiments, respectively, 
with precision \textbf{P} ($TP/(TP+FP)$) and recall \textbf{R} ($TP/(TP+FN)$) 
for each galaxy class: spiral (\textbf{S}) and elliptical (\textbf{E}). 
\textbf{\emph{F-score}} ($F_1 = 2\times{(P \times C)}/{(P+C)}$),
\emph{Overall Accuracy} (\textbf{OA} = $(TP+TN)/(TP+TN+FP+FN)$) and 
\textbf{Kappa index ($\kappa$)} for each experiment also appear in these tables.
Kappa is a statistic which measures inter-rater agreement for classification
problems. Inter-rater agreement, also known as concordance, is the degree of
agreement among raters. Thus, Kappa measures the degree of agreement beyond 
what would be expected by chance alone. This measure has a maximum value of 
1, where 1 represents total agreement; and values close to and below 0, 
indicate no agreement, or agreement was exactly the one expected by chance.

\begin{table}[!ht]
 \centering
 \begin{tabular}{|c|c|c|c|c|c|c|c|}
  \hline
  \textbf{\#Exp} & \textbf{P(S)} \% & \textbf{R(S)} \% 
    & \textbf{P(E)} \% & \textbf{R(E)} \% & \textbf{$F_1$} \% &  \textbf{OA} \% 
    & \textbf{$\kappa$} \\ \hline \hline
  \textbf{SVM1} & 98.556 & 98.684 & 82.767 & 81.383 & 0.9862 & 97.437 & 0.807 \\ \hline 
  \textbf{SVM2} & 98.601 & 98.49  & 80.631 & 81.805 & 0.9855 & 97.3   & 0.798 \\ \hline 
  \textbf{SVM3} & 98.748 & 98.454 & 79.6   & 82.864 & 0.9860 & 97.395 & 0.798 \\ \hline 
  \textbf{SVM4} & 98.818 & 98.532 & 80.06  & 83.333 & 0.9867 & 97.528 & 0.803 \\ \hline \hline
  \textbf{DT1}  & 98.863 & 98.792 & 83.586 & 84.404 & 0.9883 & 97.815 & 0.828 \\ \hline 
  \textbf{DT2}  & 98.85  & 98.707 & 82.738 & 84.37  & 0.9866 & 97.726 & 0.823 \\ \hline 
  \textbf{DT3}  & 98.682 & 98.673 & 82.416 & 82.513 & 0.9868 & 97.541 & 0.811 \\ \hline 
  \textbf{DT4}  & 98.758 & 98.703 & 83.213 & 83.819 & 0.9873 & 97.643 & 0.822 \\ \hline \hline
 \end{tabular}
 \caption{Precision (P), Recall (R) and F-score ($F_1$) for each class; 
 Overall Accuracy (OA) and $\kappa$ for each supervised experiment.}
 \label{tab:comp1}
\end{table}

\begin{table}[!ht]
 \centering
 \begin{tabular}{|c|c|c|c|c|c|c|c|}
  \hline
  \textbf{\#Exp} & \textbf{P(S)} \% & \textbf{R(S)} \%  
    & \textbf{P(E)} \% & \textbf{R(E)} \% & \textbf{$F_1$} \% &  \textbf{OA} \% 
    & \textbf{$\kappa$} \\ \hline \hline
  \textbf{K-means1} & 85,847 & 95,953 & 93,662 & 87,172 & 0.8926 & 89,806 & 0,796 \\ \hline
  \textbf{K-means2} & 85,847 & 95,953 & 93,662 & 87,172 & 0.8926 & 89,806 & 0,796 \\ \hline \hline
  \textbf{AHC} 	    & 95,171 & 84,389 & 81,921 & 94,293 & 0.8946 & 88,634 & 0,772 \\ \hline \hline
 \end{tabular}
 \caption{Precision (P), Recall (R) and F-score ($F_1$) for each class; 
 Overall Accuracy (OA) and $\kappa$ for each unsupervised experiment.}
 \label{tab:comp2}
\end{table}

In general, DTs have the best results, considering CN as the most important feature to
separate galaxies into spiral and elliptical (responsible attribute for the first decision
in all DTs). The Grid Search applied in the supervised methods optimized the OA.
Due to the unbalance in the dataset (44760 galaxies labeled as S and 3385 as E), none 
experiment reached Kappa index ($\kappa$) of 0,9, although the interval $0,8 \le \kappa 
\le 1$ is considered of excellent concordance. The recall was also affected by this 
unbalance. However, all supervised methods have over 97\% of OA.

For future works with clustering methods, we plan to build trainned models with this dataset to give 
supervision in this classification task with bigger unlabeled datasets. Also, we are studying to 
apply Generative Adversarial Networks (GAN) \cite{GAN} and other Deep Learning techniques to improve 
galaxy morphology with machine learning further.

\quad

{\bf Acknowledgments}. 
PB, RS, RRR and RRdC thank CAPES, CNPq and FAPESP for partial financial support.
RRdC acknowledges financial support from FAPESP through a grant \# 2014/11156-4.

\end{document}